# RESEARCH

**Open Access**

# Decline of COPD exacerbations in clinical trials over two decades – a systematic review and meta-regression

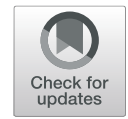

Stefan Andreas[1,2]*, Christian Röver[3], Judith Heinz[3], Sebastian Straube[4], Henrik Watz[5] and Tim Friede[3]

## Abstract

**Background:** An important goal of chronic obstructive pulmonary disease (COPD) treatment is to reduce the frequency of exacerbations. Some observations suggest a decline in exacerbation rates in clinical trials over time. A more systematic understanding would help to improve the design and interpretation of COPD trials.

**Methods:** We performed a systematic review and meta-regression of the placebo groups in published randomized controlled trials reporting exacerbations as an outcome. A Bayesian negative binomial model was developed to accommodate results that are reported in different formats; results are reported with credible intervals (CI) and posterior tail probabilities ($p_B$).

**Results:** Of 1114 studies identified by our search, 55 were ultimately included. Exacerbation rates decreased by 6.7% (95% CI (4.4, 9.0); $p_B$ < 0.001) per year, or 50% (95% CI (36, 61)) per decade. Adjusting for available study and baseline characteristics such as forced expiratory volume in 1 s ($FEV_1$) did not alter the observed trend considerably. Two subsets of studies, one using a true placebo group and the other allowing inhaled corticosteroids in the "placebo" group, also yielded consistent results.

**Conclusions:** In conclusion, this meta-regression indicates that the rate of COPD exacerbations decreased over the past two decades to a clinically relevant extent independent of important prognostic factors. This suggests that care is needed in the design of new trials or when comparing results from older trials with more recent ones. Also a considerable effect of adjunct therapy on COPD exacerbations can be assumed.

**Registration:** PROSPERO 2018 CRD4218118823.

## Background

Chronic obstructive pulmonary disease (COPD) is a major cause of death and disability worldwide, and the burden of this disorder caused by smoking will likely continue to increase despite therapeutic advances [1]. The chronic course of COPD is aggravated by disease exacerbations. Nearly 20% of exacerbations in the populations of recent large clinical trials required hospitalization [2–4]. Exacerbations reduce lung function and physical ability as well as quality of life and ability to work, and also increase the risk of death [5]. Because exacerbations impact heavily on the natural history of the disease and the utilization of health care resources, an important goal of COPD treatment is the reduction of the number of exacerbations. Thus, many phase III studies assess COPD exacerbations as their primary endpoint [6].

Interestingly, some recent trials did not demonstrate statistically significant reductions in moderate or severe COPD exacerbations despite convincing positive effects on other clinically relevant outcomes including symptoms and lung function [7, 8]. A trial's ability to demonstrate a statistically significant treatment effect on exacerbations does not only depend on the number of patients recruited and the size of the treatment effect but also on the exacerbation rate. Therefore, it is important to reliably predict the frequency of exacerbations in the planning phase of a clinical trial. Also the interpretation of epidemiologic data and metaanalysis rests on the assumption of a relatively stable exercerbation rate over time. If not due to treatment inefficacy, another reason for randomized controlled trials (RCT) failing to demonstrate a statistically significant effect for the endpoint of exacerbation rate might have been a decline in overall exacerbation rate over time, and a failure to compensate for this in the trial design. Indeed, adjunct therapies such as vaccination, better treatment of comorbidities or healthier lifestyle might have had a positive impact on exacerbation rates in general over the past years. Also, it might be speculated that the selection of patients into trials might have changed with the increased availability of more potent therapies over the years.

* Correspondence: stefan.andreas@med.uni-goettingen.de
[1]Department of Cardiology and Pneumology, University Medical Center Göttingen, Göttingen, Germany
[2]Lung Clinic Immenhausen, Immenhausen, Germany
Full list of author information is available at the end of the article

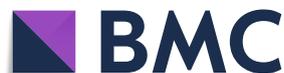

The aim of the present systematic review and meta-regression therefore was to assess whether the rate of COPD exacerbations in the placebo groups of RCTs changed according to when these trials were conducted. If such a trend was present, it would be of interest to know whether it could be explained by changing study populations in terms of relevant prognostic factors over the years.

## Methods

In conduct and reporting of this systematic review we follow the PRISMA statement [9]. This review is registered with PROSPERO (CRD42018118823, 2018).

### Literature review

In order to find relevant studies, we performed a literature search using PubMed and Cochrane CENTRAL. The search terms used were: "*(COPD OR chronic obstructive pulmonary disease OR COLD OR chronic obstructive lung disease OR 'Pulmonary Disease, Chronic Obstructive'[Mesh]) AND (double-blind OR double blind) AND exacerba\**", using the limits "*Clinical Trial*" and "*Randomized Controlled Trial*". In order to also retrieve studies that have only recently been added to the database and may not be completely indexed, we repeated the search using only the limit "*published in the last 180 days*". In addition, we considered studies that were referenced by reviews [10–12]. The date of last search was January 2nd, 2019.

The list of abstracts was independently reviewed by three authors (CR, SS, JH) using the following inclusion criteria: (i) the study needs to deal with adult COPD patients; (ii) the study has a placebo control group; (iii) the study design uses parallel group or a cross-over design; (iv) the trial is double-blind and randomized; (v) exacerbation rates are quoted or can be calculated from the data presented; (vi) at least 100 patients (intention-to-treat population) are included into the study; (vii) the study has a treatment duration of at least 12 weeks.

### Data extraction

We recorded information on patient demographics, trial eligibility criteria, exacerbation rates, and the corresponding uncertainties as well as overdispersion. Information on studies was assembled in a table and checked independently by at least two authors (CR; SS; SA; HW; JH). The reporting of rates was rather heterogeneous, partly because not all studies considered here treated exacerbation rates as a primary endpoint [13]. The preferred source of evidence on rates are explicitly quoted rate estimates, and corresponding standard errors or confidence intervals. If a rate is given along with the number of exacerbation-free patients, this pair of figures provides some evidence on the overdispersion. If only one of the numbers is given, these may still be used to fit the joint model [14]. Quoted rates or total exacerbation counts may be converted to one another, as can numbers and fractions of (non-)exacerbating patients. When exacerbation counts of differing severities are given, the numbers pertaining to "moderate to severe" exacerbation events were used. Our review focuses on data from the studies' placebo groups. Only the concurrent use of inhaled corticosteroids (ICS) or rescue medications was acceptable in order to be considered as placebo groups for purposes of our analysis. In a sensitivity analysis, we also distinguish between studies involving "true" and "ICS-placebo" groups where "true" placebo indicates that ICS were not allowed while in the "ICS-placebo" groups ICS were allowed. If ICS were allowed, this does not mean that all patients actually received ICS. Often the ICS proportion was only about 50% or commonly not reported at all. Other extracted study characteristics were: number of participants, study duration, number of patient-years, fraction of smokers, mean pack-years, mean St. George's Respiratory Questionnaire (SGRQ) score, proportion of males, and mean forced expiratory volume in 1 s ($FEV_1$). Study quality was assessed by two reviewers (SS; SA) using the Oxford Quality Scale score [15]; discrepancies in this regard were resolved by a third reviewer (TF).

### Data analysis

Modeling of the event count data here is based on the negative binomial distribution, a generalization of the Poisson distribution. The Poisson distribution is defined through a single rate parameter; for example, a patient may be assumed to experience events at an annualized rate $\lambda$. If the patient is observed over a duration $\delta$, the observed number of events is Poisson ($\delta\lambda$)-distributed with expectation $\delta\lambda$ and variance equal to the expectation. The negative binomial distribution possesses an additional *overdispersion* parameter ($\phi \geq 0$) in order to account for extra variation (or *heterogeneity*) between patients, beyond what could be accommodated by a Poisson distribution [16]. If event counts are modeled using a negative binomial distribution, the expected event count again equals the rate ($\delta\lambda$), but at $\delta\lambda (1 + \phi\delta\lambda)$ the variance may be larger, depending on the amount of overdispersion. For $\phi = 0$, the distribution again reduces to a Poisson distribution. The use of negative binomial models for parametric analyses of exacerbation counts has been advocated e.g. by [13, 17] and in the corresponding EMA guideline [18]. Published hints on the amount of overdispersion to be expected on the other hand are rare; Keene et al. [17] quote $\phi \approx 0.5$. Anzueto et al. [19] use *"an overdispersion-estimate of 1.5"* and Calverley et al. [20] use *"a correction for overdispersion of 2"*, while the actual conventions used for quantifying overdispersion may be ambiguous.

Due to the heterogeneous standards of reporting exacerbation counts, a Bayesian model was developed in order to accommodate different types of reported outcome measures in a coherent model as published previously [14]. This way, endpoints given in terms of (a) rates and standard errors (or confidence intervals), (b) rates (or total event counts) only, (c) numbers (or proportions) of patients with and without an event, or (d) a combination of total count and event-free numbers may be jointly utilized for the analysis. Analysis then primarily aims for the annual exacerbation rate, while at the same time accounting for overdispersion. Study-specific random effects are included for logarithmic rate and overdispersion, and covariables are included in the model by assuming a linear effect on the logarithmic exacerbation rate. For the $i$th study of duration $d_i$, we assume that the number of exacerbations follows a negative binomial distribution with rate $d_i\lambda_i$ and overdispersion $\phi_i$, where $\log (\lambda_i) \sim \text{Normal}(\beta_0 + \beta_1 x, \sigma_\lambda^2)$ and $\log (\phi_i) \sim \text{Normal} (\mu_\phi, \sigma_\phi^2)$. Here $x$ is our covariable (e.g., publication year, in years since 2000). The parameters $\beta_0$ and $\beta_1$ here are the intercept and slope of the regression line, respectively, and $\sigma_\lambda^2$ and $\sigma_\phi^2$ are the heterogeneity parameters for rate and overdispersion, respectively. Further covariables then are added analogously. In order to fit the Bayesian models, we used the following prior distributions for the parameters: a uniform distribution on the interval log (0.001) to log (1000) for the intercept, i.e., $\beta_0 \sim \text{Uniform} (\log (0.001), \log (1000))$; a normal distribution with mean 0 and variance 100 for the slope, i.e., $\beta_1 \sim \text{Normal} (0, 10^2)$; a half-normal distribution with scale 1.0 for the heterogeneity of the rate, i.e., $\sigma_\lambda \sim$ half-Normal (1.0); a uniform distribution on the interval log (0.0001 to log (10000) for the mean overdispersion, i.e., $\mu_\phi \sim$ Uniform (log (0.0001), log (10000)); and a half-normal distribution with scale 1.0 for the heterogeneity of the overdispersion, i.e., $\sigma_\phi \sim$ halfNormal (1.0). The models are fitted via Markov chain Monte Carlo (MCMC) methods using JAGS and the "rjags" R package [21]. Estimates are quoted in terms of posterior medians and central 95% credible intervals (CIs). For the Bayesian analyses, we quote two-sided posterior tail probabilities ($p_B$) for the regression parameters instead of frequentist (two-sided) *p*-values.

The model is primarily used to fit a time trend in exacerbation rates. As an additional analysis, similar to the strategy adopted by Steinvorth et al. [22], we investigate whether other covariables exhibit a correlation with the publication year, which might help establishing a causal connection. If a correlation is found to be statistically different from zero, the corresponding variable is considered as an additional explanatory variable in the regression model. Since treatment effects are not in the focus of the present study, we do not expect systematic biases in the reporting of results.

## Results
### Included studies and their characteristics
The literature review resulted in a total of 1114 distinct studies; of these, 283 full texts were screened for eligibility, and 55 were eventually included in the quantitative analysis (see also the PRISMA flow chart in Fig. 1). The studies included 14,065 placebo patients, covering 10,491 patient-years. Study durations ranged from 12 weeks to 3 years, with a median of half a year. The median number of placebo patients was about 200, with a range in group size from 43 up to 1500. The included studies' characteristics are summarized in Table 1. The studies were of overall good quality; the minimum Oxford Quality Score of 2/5 (since randomization and blinding were inclusion criteria) was achieved by one study only, and a majority of 64% of the studies reached scores of 4 or 5/5.

### Temporal trends in exacerbation rates
Data on exacerbations were reported in various formats. Three studies provided data in terms of a rate estimate and standard error, 14 studies provided a total exacerbation count and the number of exacerbation-free patients, 9 studies quoted only the total count, and 29 studies gave only the proportion of non-exacerbating patients. The effect of publication year on exacerbation rate is estimated to decrease (95% CI) by 6.7% (4.4, 9.0) per year, or 50% (36, 61) per decade. Figure 2 illustrates the estimated annualized exacerbation rates. Other parameter estimates are shown in Table 2. Considering the two subsets of studies using "true placebos" or "ICS-placebos" yields very similar results; for these we calculate an estimated decrease (95% CI) of 7.5% (2.6, 12.4) per year (54% (23, 73) per decade) and 6.7% (3.7, 9.6) per year (50% (33, 64) per decade), respectively. The corresponding trends with CIs are illustrated in Fig. 2 and the estimates are shown in Table 3 and Fig. 4.

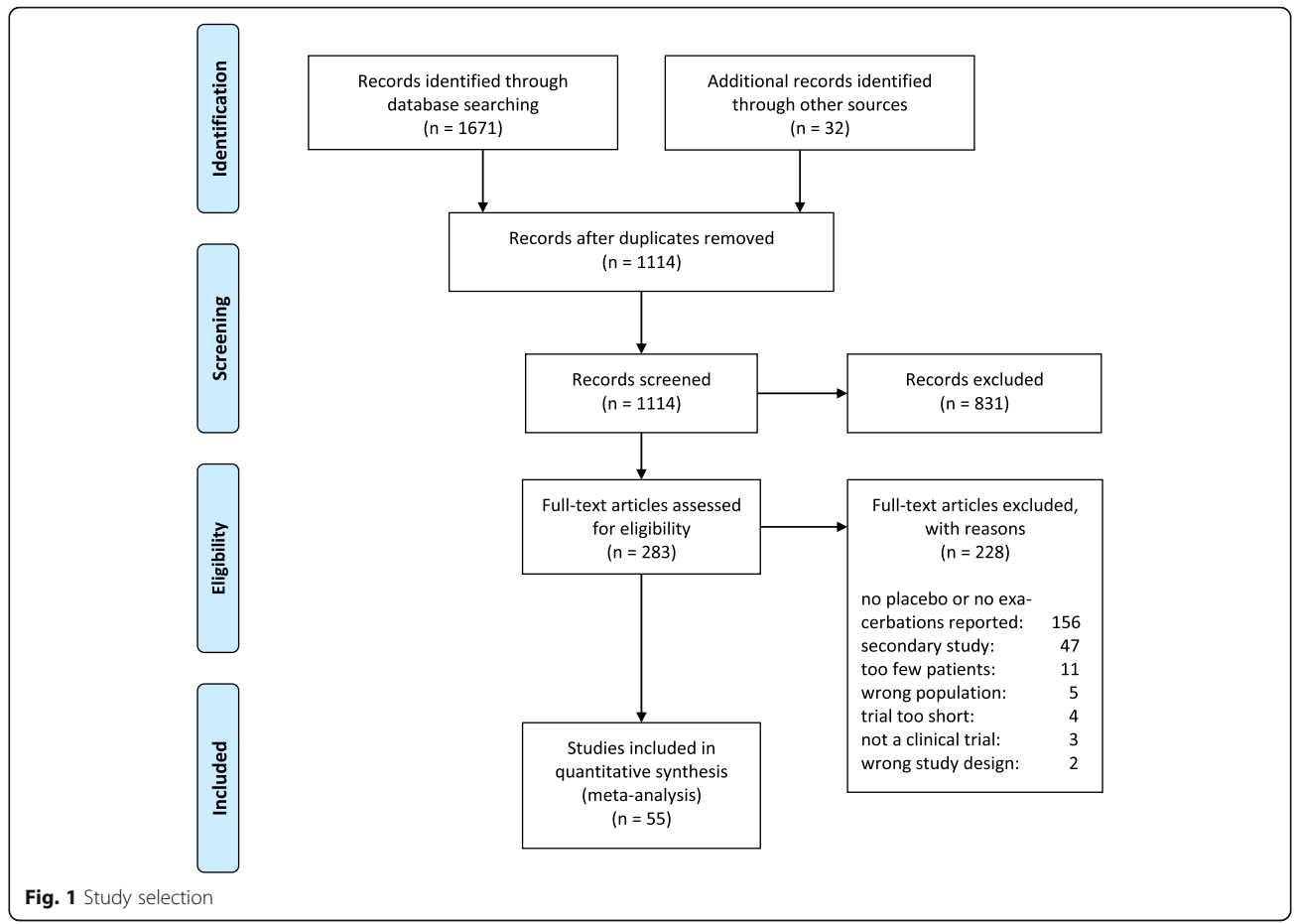

**Fig. 1** Study selection

### Trends in study and patient characteristics

Among the remaining available characteristics of the studies and their populations (i.e., number of patients, study duration, number of patient-years, fraction of smokers, mean pack-years, mean St. George's Respiratory Questionnaire (SGRQ), proportion of males, mean

**Table 1** characteristics of the included studies, overall and in the two subgroups ("true" and ICS-placebos)

|  | All studies | | | "True" placebos | | | ICS-placebos | | |
|---|---|---|---|---|---|---|---|---|---|
|  | N | Median | Range | N | Median | Range | N | Median | Range |
| Patients | 55 | 207 | (43–1524) | 21 | 219 | (56–1524) | 34 | 197.5 | (43–753) |
| Study duration (yr) | 55 | 0.4808 | (0.2308–3) | 21 | 0.4615 | (0.2308–3) | 34 | 0.5 | (0.2308–1) |
| Mean followup (yr) | 55 | 0.4327 | (0.1986–2.2596) | 21 | 0.4296 | (0.2184–2.2596) | 34 | 0.4481 | (0.1987–0.9024) |
| Mean age (yr) | 54 | 63.9 | (58.8–68.6) | 21 | 64.9 | (58.8–68.2) | 33 | 63.5 | (60–68.6) |
| Males (%) | 55 | 74.4 | (32.9–100) | 21 | 75 | (32.9–94.2) | 34 | 72.9 | (51.6–100) |
| Smokers (%) | 45 | 43.4 | (16.9–63) | 19 | 39.7 | (23–63) | 26 | 44.65 | (16.9–56) |
| Mean pack-years | 44 | 44 | (29.4–60.2) | 18 | 43.75 | (31.6–56.1) | 26 | 44.3 | (29.4–60.2) |
| Mean FEV-1 | 50 | 50.25 | (36–73.2) | 20 | 47.8 | (36–73.2) | 30 | 52.6 | (40.3–71.5) |
| Mean SGRQ | 28 | 46.52 | (33.1–55.6) | 10 | 47.55 | (42.59–55.6) | 18 | 46.07 | (33.1–52) |

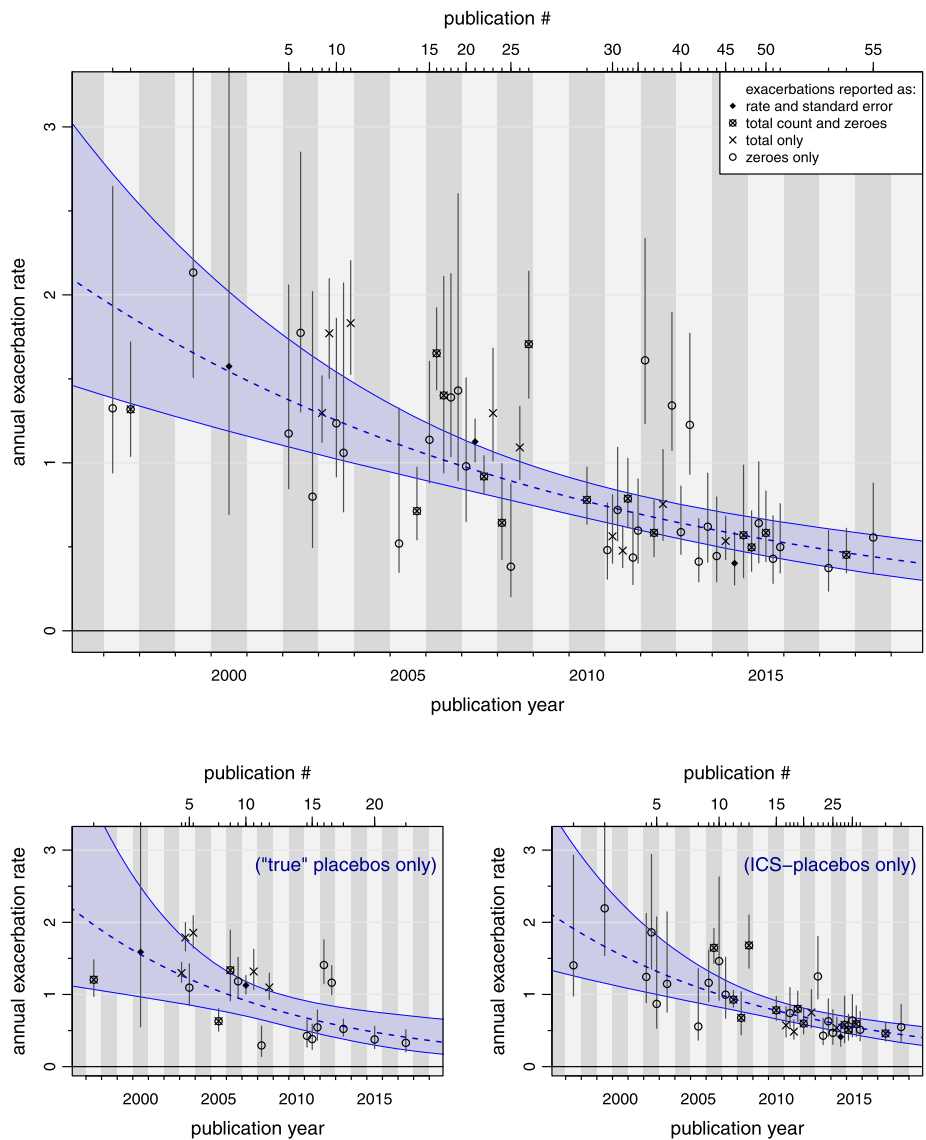

**Fig. 2** Time trend in annualized exacerbation rates based on the Bayesian regression analysis (without adjustment for study or baseline characteristics). The dashed line shows the estimated rate over time along with a 95% credible region. The points and vertical whiskers indicate each individual study's estimated rate along with a 95% credible interval. The shown ("shrinkage") estimates are based not only on the corresponding study's provided data (which in some cases may not be sufficient to derive a rate estimate), but these are supported by the complete data set. Different types of data sources are indicated by different symbols (some references directly provide rate estimates along with standard errors, some report the total number of exacerbations and the number of exacerbation-free patients, and some only one of the two). Within each publication year, estimates are evenly distributed so that they are identifiable by their respective index (see axis at top and Additional file 1: Table S1). Inserted below are two smaller figures illustrating the time trend for the two subsets of "true" placebos (21 studies) and ICS-placebos (34 studies)

age, mean $FEV_1$ and Oxford Quality Scale score), only the mean SGRQ score and $FEV_1$ exhibited statistically significant correlations with publication year. Correlation coefficients (95% CI; p-value) are at $r = -0.50$ ((− 0.74, − 0.16); $p = 0.0062$) and $r = +0.36$ ((0.09, 0.58); $p = 0.010$), respectively, in both cases indicating a decrease in (mean) severity among included patients over time. The changes in mean SGRQ scores and $FEV_1$ values over time are also shown in Fig. 3.

### Explaining the temporal trend in exacerbation rates

The SGRQ was available for 28 studies. The inclusion of mean SGRQ score as a covariable into the meta-analysis model led to a meta-regression resulting in a similar estimate for the annual decrease in exacerbation rate. The parameter estimates of the meta-regression are shown in Table 4. The effect of SGRQ is near neutral, suggesting 0.1% decrease in rate per 1 point increase in SGRQ score, and the CI is wide

**Table 2** Parameter estimates from regression analysis (all 55 studies). The parameters originally refer to the rate on the logarithmic scale; the effects may also be expressed in terms of a corresponding percentage change (last column). Bayesian posterior tail probabilities ($p_B$) are also provided for the regression coefficients instead of frequentist two-sided *p*-values

| Parameter | Estimate | Implied annual percentage change | $P_b$ |
|---|---|---|---|
| Rate intercept $\beta_0$ | 0.434 (0.172, 0.702) | | |
| Rate slope $\beta_1$ (publication year) | −0.070 (−0.095, −0.045) | −6.719 (−9.029, −4.384) | < 0.001 |
| Rate random effect $\sigma_\lambda$ | 0.409 (0.310, 0.532) | | |
| Overdispersion mean $\mu_\varphi$ | −0.092 (−0.913, 0.384) | | |
| Overdispersion random effect $\sigma_\varphi$ | 0.709 (0.277, 1.559) | | |

(from − 4.0 to + 3.9%) and includes zero, i.e. no significant change.

Consideration of mean baseline $FEV_1$ also led to consistent results. Mean $FEV_1$ was available for 50 studies; adjustment for $FEV_1$ led again to a similar, although slightly reduced, estimate of the annual decrease. The estimated effect of $FEV_1$ amounts to a 2.8% reduction in rate for each percentage point increase in $FEV_1$, with a CI ranging from 1.1 to 4.5%. All parameter estimates are shown in Table 4, and the estimates of the annual reduction in exacerbation rate resulting from the different models discussed above are illustrated side-by-side in Fig. 4.

In addition, we considered a model adjusting the publication year effect for mean baseline SGRQ as well as mean baseline $FEV_1$. Only 25 studies were available providing information on SGRQ as well as $FEV_1$. With this reduced number of studies and additional parameters, we expect more uncertainty in the parameter estimates. Table 4 shows the results with the correspondingly wider CIs, which are still consistent with the previous analyses.

## Discussion

Considering placebo arms of RCTs, we found a drop in COPD exacerbations of 50% in a decade, which is a substantial effect as generally reductions of at least 20% in exacerbation rates compared to placebo are considered clinically relevant [23]. Known predictors of exacerbation rate such as $FEV_1$, age, sex, symptoms, smoking status and smoking history have little effect on this finding. To the best of our knowledge this is the first meta-regression analysis of RCT investigating a time trend in the frequency of COPD exacerbations.

Previous epidemiologic data are in line with our findings. A Canadian inception cohort of patients hospitalized for COPD during 1990 to 2005 found that the time from the first to the second severe exacerbation increased after the year 2000 compared with before 2000 [5]. Similar time trends were reported from Spain [24], hinting at a general trend in the Western world.

Recent large-scale RCTs suggest that it is more difficult to demonstrate a treatment effect in relatively affluent regions such as North America and Western Europe as compared to Africa, Asia and Eastern Europe where COPD treatment might be less advanced and comparable to treatment in Western Europe some time ago [3]. One reason are the lower exacerbation rates observed in some regions compared to others [3], which is in line with our finding that advances in health

**Table 3** Parameter estimates (analogous to Table 2) for the regression analyses for the subgroups of "true placebos" and "ICS-placebos"

| Parameter | Estimate | Implied annual percentage change | $P_b$ |
|---|---|---|---|
| True placebos (21 studies): | | | |
| Rate intercept $\beta_0$ | 0.440 (− 0.034, 0.912) | | |
| Rate slope $\beta_1$ (publication year) | −0.079 (− 0.132, − 0.026) | − 7.553 (− 12.356, − 2.606) | 0.006 |
| Rate random effect $\sigma_\lambda$ | 0.517 (0.356, 0.787) | | |
| Overdispersion mean $\mu_\varphi$ | −5.139 (− 9.011, − 0.708) | | |
| Overdispersion random effect $\sigma_\varphi$ | 0.667 (0.031, 2.263) | | |
| ICS-placebos (34 studies): | | | |
| Rate intercept $\beta_0$ | 0.440 (−0.034, 0.912) | | |
| Rate slope $\beta_1$ (publication year) | −0.079 (− 0.132, − 0.026) | −7.553 (− 12.356, − 2.606) | < 0.001 |
| Rate random effect $\sigma_\lambda$ | 0.517 (0.356, 0.787) | | |
| Overdispersion mean $\mu_\varphi$ | −5.139 (− 9.011, − 0.708) | | |
| Overdispersion random effect $\sigma_\varphi$ | 0.667 (0.031, 2.263) | | |

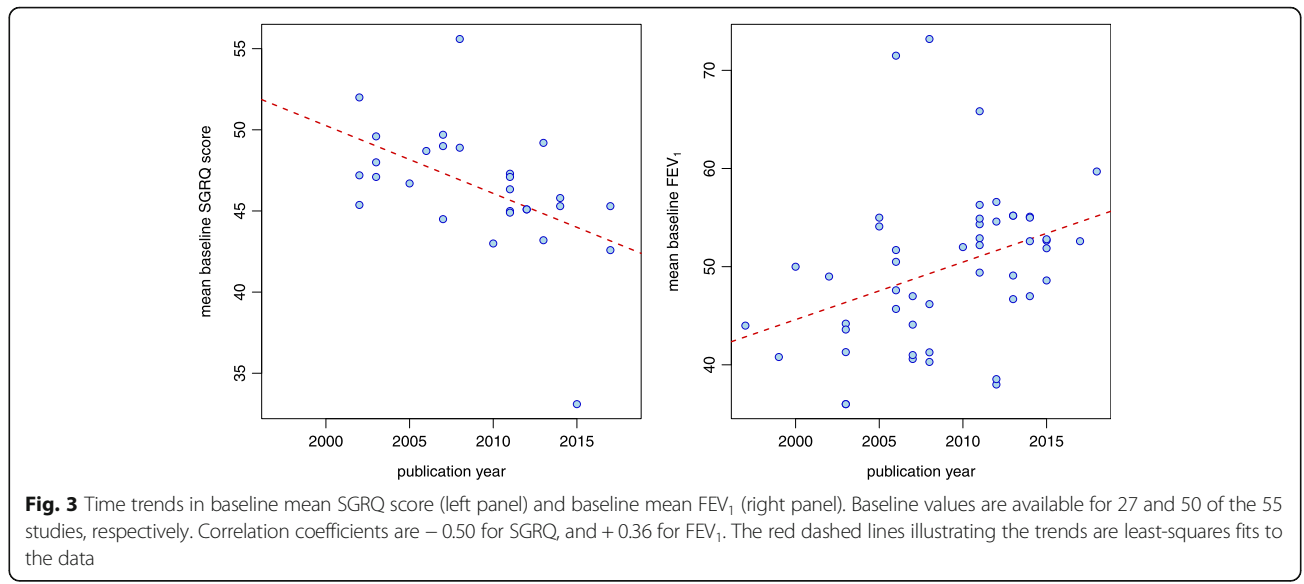

**Fig. 3** Time trends in baseline mean SGRQ score (left panel) and baseline mean $FEV_1$ (right panel). Baseline values are available for 27 and 50 of the 55 studies, respectively. Correlation coefficients are − 0.50 for SGRQ, and + 0.36 for $FEV_1$. The red dashed lines illustrating the trends are least-squares fits to the data

care over time came with reduced exacerbation rates in the trials.

We speculate that adjunct treatments in the placebo (and verum) group improved over time. Vaccination, exercise or as-needed short acting bronchodilator treatments might have been used more frequently. Better treatment of comorbidities such as cardiovascular disease [25], better ambient air quality and a healthier lifestyle might conceivably also have contributed [26].

**Table 4** Parameter estimates (analogous to Table 2) for regression analyses adjusting for SGRQ score or $FEV_1$

| Parameter | Estimate | Implied percentage change | $p_B$ |
|---|---|---|---|
| Adjusting for SGRQ (28 studies): | | | |
| Rate intercept $\beta_0$ | 0.760 (−1.301, 2.804) | | |
| Rate slope $\beta_1$ (publication year) | −0.096 (− 0.130, − 0.058) | −9.114 (− 12.205, − 5.644) | < 0.001 |
| Rate slope $\beta_2$ (SGRQ score) | −0.001 (− 0.041, 0.038) | −0.093 (− 4.015, 3.923) | 0.961 |
| Rate random effect $\sigma_\lambda$ | 0.247 (0.145, 0.415) | | |
| Overdispersion mean $\mu_\varphi$ | 0.263 (−0.645, 0.917) | | |
| Overdispersion random effect $\sigma_\varphi$ | 0.954 (0.414, 1.723) | | |
| adjusting for $FEV_1$ (50 studies): | | | |
| Rate intercept $\beta_0$ | 1.674 (0.876, 2.482) | | |
| Rate slope $\beta_1$ (publication year) | −0.050 (−0.079, −0.022) | −4.912 (− 7.598, − 2.172) | 0.001 |
| Rate slope $\beta_2$ ($FEV_1$) | − 0.028 (− 0.046, − 0.011) | −2.810 (− 4.517, −1.093) | 0.002 |
| Rate random effect $\sigma_\lambda$ | 0.374 (0.279, 0.498) | | |
| Overdispersion mean $\mu_\varphi$ | − 0.150 (−1.008, 0.340) | | |
| Overdispersion random effect $\sigma_\varphi$ | 0.726 (0.299, 1.508) | | |
| Adjusting for SGRQ and $FEV_1$ (25 studies): | | | |
| Rate intercept $\beta_0$ | 2.664 (−0.261, 5.583) | | |
| Rate slope $\beta_1$ (publication year) | −0.045 (− 0.102, 0.008) | −4.404 (−9.737, 0.816) | 0.094 |
| Rate slope $\beta_2$ (SGRQ score) | −0.009 (− 0.051, 0.035) | −0.849 (− 4.963, 3.610) | 0.690 |
| Rate slope $\beta_3$ ($FEV_1$) | −0.042 (− 0.079, − 0.004) | −4.131 (− 7.614, − 0.369) | 0.032 |
| Rate random effect $\sigma_\lambda$ | 0.269 (0.161, 0.426) | | |
| Overdispersion mean $\mu_\varphi$ | 0.123 (− 0.777, 0.777) | | |
| Overdispersion random effect $\sigma_\varphi$ | 0.744 (0.273, 1.598) | | |

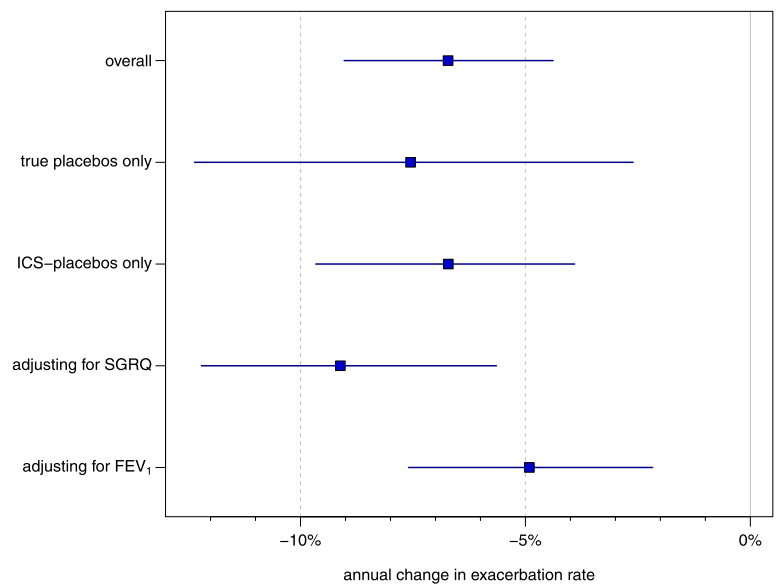

Fig. 4 Estimated annual reduction in exacerbation rates (see also Tables 2,3,4). In the "overall" model, the estimated decrease is at 6.3% (95% CI (3.9, 8.7)) per calender year. The estimates based on the two disjoint subgroups of placebo controls are very similar, and adjusting for baseline mean SGRQ score or $FEV_1$ also leads to consistent effects

Several baseline characteristics predictive of exacerbation rate were unknown and thus not reported at the beginning of COPD RCTs two decades ago. Only more recent trials identified and reported further predictors of exacerbation such as the 6 min walk test, physical activity, previous exacerbations or small airway abnormality on CT [27–29]. Similarly, biomarkers indicative of inflammation such as C-reactive protein, fibrinogen, white blood cell count or eosinophil count [27, 30, 31] were only reported in more recent studies. In the present meta-analysis we have only included well known predictors of exacerbation rate that were reported in a sufficient number of studies (i.e. fraction of smokers, mean pack-years, mean SGRQ score, proportion of males, mean age, and mean $FEV_1$). Only SGRQ score and $FEV_1$ exhibited statistically significant correlations with publication year.

In less than 50% of studies health related quality of life as evaluated by the SGRQ was reported. SGRQ values have decreased over the years indicating less severe disease (Fig. 3). However, the observed time trend in exacerbation rate was little affected by this finding (Fig. 4).

$FEV_1$ in % predicted was reported in nearly all studies and increased over time (Fig. 3). Again the observed time trend was little affected by $FEV_1$ (Fig. 4). In our analysis we used $FEV_1$ in % of the predicted value and not the absolute values. Previous publications did not show any difference between both measures concerning exacerbations or parameters describing disease severity [27].

Due to the ambiguity in the definition of what exactly constitutes an exacerbation, studies usually may not be directly comparable with one another, and we expect some heterogeneity in the findings from different studies. Effing et al. demonstrated that the use of different conventions may make substantial differences in the magnitude and significance of the findings [32]. Although all definitions contain some subjective judgements, reproducibility and validity are nevertheless generally high. Problems may arise, however, if studies using different definitions are meta-analyzed [10, 33, 34]. Our approach was to use data for moderate to severe exacerbations, if more than one category of exacerbations was reported and we should thereby have minimized the effect of differences in the definitions of exacerbations used. Furthermore, we believe it is unlikely that our findings can be explained by a change in the definition of exacerbation over time.

Although exacerbations are commonly considered as primary outcome in confirmatory trials in COPD, their reporting is not consistent across trials. While most large RCT report annualized rates, many studies give only the number of exacerbation-free patients, or, alternatively, the numbers of patients with no, one, two or three exacerbations. In the present analysis we have dealt with this problem by modelling the data jointly using a negative binomial model with random effects for rate and overdispersion as reported previously [14].

In our analysis concomitant ICS monotherapy had little effect on the position or slope of the time trend in exacerbation rates (Fig. 2) compared to true placebo arms. It should be noted, however, that a substantial portion of the "ICS-placebo" group was not actually treated with ICS.

The findings of our meta-regression have important implications for the planning of future clinical trials and the comparison of treatment effects of earlier studies: Since sample size calculations cannot rely on the event rate being stable over time, an adaptive trial design might be more appropriate [35]. Indeed, the US Food and Drug Administration guidance on adaptive designs recommends that sample size adjustment using blinded methods to maintain desired study power should be considered [21].

Network meta-analyses are commonly used to indirectly compare treatments that have not been compared directly in a RCT [36]. However, the decrease in exacerbation rate over time as observed in our analysis highlights the difficulties with such an approach. For instance a network meta-analysis on inhaled drugs to reduce COPD exacerbations included studies over a 15 year period [10]. Another meta-analysis on COPD exacerbations encompassed a time span of 7 years [13]. Other meta-analyses or databases of routinely collected healthcare data integrated studies over a period of 10 [37], 13 [38], and 17 years [39].

Relapsing–remitting multiple sclerosis is another disease with an infrequent but clinical relevant endpoint evaluated in clinical trials. Similar to our work in COPD, a decline in event rates over the years was observed [22] that complicates the interpretation of differences in meta-analysis [40].

### Limitations and strengths

Although we took great care to distinguish different definitions of exacerbations, some idiosyncrasies between studies cannot be classified and analyzed. As discussed above, we were able to analyze only commonly reported predictors of exacerbations. Thus we cannot exclude that unidentified predictors changed over time. Given the comparatively small effect of well-established predictors of exacerbations such as SGRQ and $FEV_1$, however, we consider it unlikely that this would be a significant limitation.

The main strength of our study lies in the long period of time evaluated and the consistent findings following adjustment for a number of well known confounders including age, smoking status, baseline symptoms and lung function. Furthermore, we consider it a strength that the statistical methods used allow for the consideration of heterogeneous ways of reporting exacerbations.

### Conclusions

This systematic review with meta-regression analysis indicates that the rate of COPD exacerbations in placebo groups of clinical trials decreased over two decades to a clinical relevant extent, independent of important confounders. This finding is consistent with the presumption of a substantial effect of adjunct treatments on exacerbations. Furthermore, care is needed when planning studies or comparing older with more recent studies using exacerbations as an endpoint.

### Additional files

**Additional file 1:** Table S1. The list of references ordered by publication year, and alphabetically within each year. Publication numbers correspond to numbers shown at the top of Fig. 2. (PDF 476 kb)

**Additional file 2:** containing the data entering the analyses. (CSV 5 kb)

**Additional file 3:** The PRISMA checklist. (PDF 671 kb)

#### Abbreviations
CI: credible interval; COPD: chronic obstructive pulmonary disease; $FEV_1$: forced expiratory volume in 1 s; ICS: inhaled corticosteroid; RCT: randomized controlled trial; SGRQ: St. George's respiratory questionnaire

#### Acknowledgements
Not applicable.

#### Authors' contributions
TF and SA conceived of the study. CR, SA, SS, HW, JH, TF contributed to study selection. CR, JH and TF analyzed the data. AS and TF drafted the manuscript and all authors revised the manuscript.All authors read and approve the final manuscript.

#### Funding
Part of the work was funded by an unrestricted contribution to Stefan Andreas (Oskar und Helene Medizinpreis 2010).

#### Availability of data and materials
All data generated and/or analysed during the study are included in the published article and its Additional files 1, 2 and 3.

#### Ethics approval and consent to participate
Not applicable.

#### Consent for publication
Not applicable.

#### Competing interests
Dr. Andreas reports other from Stiftung Oskar-Helene-Heim, during the conduct of the study; grants and personal fees from Boehringer Ing, grants from Pfizer, personal fees from Novartis, personal fees from Astra Zeneca, personal fees from GSK, personal fees from Chiesi, personal fees from Merini, outside the submitted work. Dr. Röver has nothing to disclose. Dr. Heinz has nothing to disclose. Dr. Straube reports grants and personal fees from Workers' Compensation Board of Alberta, personal fees from Canadian Board of Occupational Medicine, outside the submitted work. Dr. Watz reports personal fees from Novartis, personal fees from Bayer, personal fees from AstraZeneca, personal fees from Roche, personal fees from Boehringer Ingelheim, personal fees from GSK, personal fees from BerlinChemie, personal fees from Chiesi, outside the submitted work. Dr. Friede reports personal fees from Novartis, personal fees from Bayer, personal fees from AstraZeneca, personal fees from Janssen, personal fees from SGS, personal fees from Roche, personal fees from Mediconomics, personal fees from Boehringer Ingelheim, personal fees from Daiichi-Sankyo, personal fees from Galapagos, personal fees from Penumbra, personal fees from Parexel, outside the submitted work.

#### Author details
[1]Department of Cardiology and Pneumology, University Medical Center Göttingen, Göttingen, Germany. [2]Lung Clinic Immenhausen, Immenhausen,

Germany. ³Department of Medical Statistics, University Medical Center Göttingen, Göttingen, Germany. ⁴Division of Preventive Medicine, Department of Medicine, University of Alberta, Edmonton, Canada. ⁵Pulmonary Research Institute at LungenClinic Grosshansdorf, Airway Research Center North (ARCN), German Center for Lung Research (DZL), Grosshansdorf, Heidelberg, Germany.

## Publisher's Note

Springer Nature remains neutral with regard to jurisdictional claims in published maps and institutional affiliations.